\documentclass{ws-procs9x6}
\usepackage{amsmath}
\usepackage{amssymb}
\usepackage{amsfonts}
\usepackage{bm}
\begin{document}

\title{The Skyrme Model Revisited: \\ 
An Effective Theory Approach and Application to the
Pentaquarks\footnote{\uppercase{T}his talk is a preliminary version of
the work in the collaboration with 
\uppercase{Y}.~\uppercase{M}itsunari and
\uppercase{N}.~\uppercase{Y}amashita, hep-ph/0410145.}}

\author{Koji Harada\footnote{\uppercase{W}ork partially supported by
\uppercase{G}rant-in-\uppercase{A}id for \uppercase{S}cientific
\uppercase{R}esearch on \uppercase{P}riority \uppercase{A}rea,
\uppercase{N}umber of \uppercase{A}rea 763, ``\uppercase{D}ynamics of
\uppercase{S}trings and \uppercase{F}ields,'' from the
\uppercase{M}inistry of \uppercase{E}ducation, \uppercase{C}ulture,
\uppercase{S}ports, \uppercase{S}cience and \uppercase{T}echnology,
\uppercase{J}apan.}}  

\address{Department of Physics, Kyushu University\\ 
Fukuoka 810-8581 Japan 
\\ E-mail: koji1scp@mbox.nc.kyushu-u.ac.jp}

\maketitle

\abstracts{ The Skyrme model is reconsidered from an effective theory
point of view. Starting with the most general Lagrangian up to including
terms of order $p^4$, $N_c$ and $\delta m^2$ ($\delta m\equiv m_s-m$),
we obtain new interactions, which have never been discussed in the
literature.  We obtain the parameter set best fitted to the low-lying
baryon masses by taking into account the representation mixing up to
$\bm{27}$. A prediction for the mainly anti-decuplet excited nucleon
$N'$ and $\Sigma'$ is given. }

\section{Introduction and Summary}

The narrowness of the newly discovered exotic baryonic resonance
$\Theta^{+}$
\cite{Nakano:2003qx,Barmin:2003vv,Stepanyan:2003qr,Barth:2003es} has
been a mystery. The direct experimental upper bound is
$\Gamma_{\Theta}<9$ MeV, while some re-examinations
\cite{Nussinov:2003ex,Arndt:2003xz,Workman:2004yd,Cahn:2003wq} of older
data suggest $\Gamma_{\Theta}<1$ MeV. At this moment, it is not very
clear what makes the width so narrow.

Interestingly, the mass and its narrow width had been predicted by
Diakonov, Petrov, and Polyakov\cite{Diakonov:1997mm}. Compare their
predicted values, $M_\Theta=1530$ MeV and $\Gamma=15$ MeV (or 30
MeV\cite{Jaffe:2004qj,Diakonov:2004ai,Jaffe:2004dc}), with the
experimental ones\cite{Eidelman:2004wy}, $M_\Theta=1539.2\pm 1.6$ MeV
and $\Gamma=0.9\pm 0.3$ MeV. It is astonishing! What allows the authors
to predict these numbers? It deserves a serious look.

Their predictions are based on the ``chiral quark-soliton
model\cite{Diakonov:2000pa},'' ($\chi$QSM) which may be regarded as a
version of the Skyrme model\cite{Skyrme:1961vq} with specific symmetry
breaking interactions\footnote{The $\chi$QSM has its own scenario based
on chiral symmetry breaking due to instantons. But for our purpose, it
is useful to regard it as a Skyrme model.},
\begin{equation}
 \alpha D^{(8)}_{88}+\beta Y 
  +\frac{\gamma}{\sqrt{3}}\sum_{i=1}^3 D^{(8)}_{8i}J_i,
  \label{DPP}
\end{equation}
where $ D^{(8)}_{\alpha\beta}(A) = \frac{1}{2} {\rm Tr}\left(A^\dagger
\lambda_\alpha A \lambda_\beta\right)$, $Y$ is the hypercharge operator,
and $J_i$ is the spin operator. Is this a general form of the symmetry
breaking? Is it possible to justify it without following their long
way, just by relying on a more general argument? What is the most general
Skyrme model? Is it possible to have a ``model-independent'' Skyrme
model? This is our basic motivation.

A long time ago, Witten\cite{Witten:1979kh} showed that a soliton
picture of baryons emerges in the large-$N_c$ limit\cite{'tHooft:1973jz}
of QCD. If the large-$N_c$ QCD has a close resemblance to the real QCD,
we may consider an effective theory (not just a model) of baryons based
on the soliton picture, which may be called as the ``Skyrme-Witten
large-$N_c$ effective theory.'' The question is in which theory the
soliton appears.

A natural candidate seems the chiral perturbation theory ($\chi$PT),
because it represents a low-energy QCD at least in the meson sector.
Note that it is different from the conventional Skyrme model, which
contains only a few interactions. We have now an infinite number of
terms. We have to systematically treat these infinitely many
interactions. Because we are interested in the low-energy region, we
only keep the terms up to including $\mathcal{O}(p^4)$, where $p$ stands
for a typical energy/momentum scale. Because we consider the baryons as
solitons, we keep only the leading order terms in $N_c$. In this way, we
arrive at the starting Lagrangian.

We quantize the soliton by the collective coordinate quantization, where
only the ``rotational'' modes are treated as dynamical. The resulting
Hamiltonian contains a set of new interactions, which have never been
considered in the literature. We calculate the matrix elements by using
the orthogonality of the irreducible representation of $SU(3)$ and the
Clebsch-Gordan coefficients. By using these matrix elements, we
calculate the baryon masses in perturbation theory with respect to the
symmetry breaking parameter $\delta m\equiv m_s-m$, where $m_s$ is the
strange quark mass and $m$ stands for the mass for the up and down
quarks. We ignore the isospin breaking in this work.

The calculated masses contain undetermined parameters. In the
conventional Skyrme model calculations, they are determined by the
profile function of the soliton and the $\chi$PT theory parameters. In
our effective theory approach, however, they are just parameters to be
fitted, because there are infinitely many contributions from higher
order terms which we cannot calculate. After fitting the parameters, we
make predictions.

\section{The Hamiltonian}

Let us start with the $SU_f(3)$ $\chi$PT action which includes the terms
up to $\mathcal{O}(p^4)$\cite{Gasser:1984gg},
\begin{eqnarray}
 S^{\chi{\rm PT}}&=&\frac{F_0^2}{16}\int d^4x 
  {\rm Tr}\left(\partial_\mu U \partial^\mu U^\dagger\right) 
  +\frac{F_0^2 B_0}{8}\int d^4x
  {\rm Tr}\left(\mathcal{M}^\dagger U+\mathcal{M}U^\dagger\right) \nonumber \\
 &&{}+N_c\Gamma[U] +\int d^4x \mathcal{L}_4,
% &&{}+L_1\!\int \!\!d^4x
%  \left[
%   {\rm Tr}\left(\partial_\mu U\partial^\mu U^\dagger\right)
%  \right]^2 \!\!
%  +L_2 \!\int \!\!d^4x
%  {\rm Tr}\left(\partial_\mu U^\dagger \partial_\nu U\right)\!
%  {\rm Tr}\left(\partial^\mu U^\dagger \partial^\nu U\right) \nonumber \\
% &&{}+L_3\int \!\!d^4x
%  {\rm Tr}\left(\partial_\mu U^\dagger \partial^\mu U 
%	   \partial_\nu U^\dagger \partial^\nu U\right) \nonumber \\
% &&{}+L_4B_0\int d^4x
%  {\rm Tr}\left(\partial_\mu U^\dagger \partial^\mu U\right)
%  {\rm Tr}\left(\mathcal{M}^\dagger U+\mathcal{M}U^\dagger\right) \nonumber \\
% &&{}+L_5B_0\int d^4x
%  {\rm Tr}
%  \left[
%   \partial_\mu U^\dagger \partial^\mu U
%   \left(\mathcal{M}^\dagger U+U^\dagger \mathcal{M}\right)
%  \right] \nonumber \\
% &&{}+L_6B_0^2\int d^4x
%  \left[
%   {\rm Tr}\left(\mathcal{M}^\dagger U+\mathcal{M}U^\dagger\right)
%  \right]^2
%  \nonumber \\
% &&{}+L_7B_0^2\int d^4x
%  \left[{\rm Tr}\left(\mathcal{M}^\dagger U-\mathcal{M}U^\dagger\right)\right]^2
%  \nonumber \\
% &&{}+L_8B_0^2\int d^4x
%  {\rm Tr}\left(\mathcal{M}^\dagger U\mathcal{M}^\dagger U
%	   +\mathcal{M}U^\dagger\mathcal{M}U^\dagger\right), 
\end{eqnarray}
where $ \mathcal{L}_4=\sum_{i=1}^8L_i\mathcal{O}^i$ is the terms of
$\mathcal{O}(p^4)$, $\mathcal{M}$ is the quark mass matrix, $\mathcal{M}={\rm
diag}(m,m,m_s)$, and $\Gamma$ is the WZW term\cite{Wess:1971yu,Witten:1983tw}.
%\begin{equation}
% \mathcal{M}=\left(
% \begin{array}{ccc}
%  m&0&0 \\
%  0&m&0 \\
%  0&0&m_s
% \end{array}\right).
%\end{equation}

The large-$N_c$ dependence of these low-energy coefficients are
known\cite{Gasser:1984gg,Peris:1994dh}:
\begin{eqnarray}
 B_0,2L_1-L_2,L_4,L_6,L_7 &\cdots& \mathcal{O}(N_c^0), \\
 F_0^2,L_2,L_3,L_5,L_8 &\cdots& \mathcal{O}(N_c^1).
\end{eqnarray}
As explained in the previous section, we keep only the terms of order
$N_c$. Furthermore, we assume that the constants $L_1$, $L_2$ and $L_3$
have the ratio,
\begin{equation}
 L_1:L_2:L_3=1:2:-6,
\end{equation}
which is consistent with the experimental values, $L_1=0.4\pm 0.3$,
$2L_1-L_2=-0.6\pm 0.5$, and $L_3=-3.5\pm 1.1$ (times
$10^{-3}$)\cite{Pich:1995bw}. It enables us to write the three terms in
a single expression,
\begin{equation}
  \sum_{i=1}^3L_i\mathcal{O}^i
   =\frac{1}{32e^2}{\rm Tr}
  \left(
   \left[
    U^\dagger \partial_\mu U,U^\dagger\partial_\nu U
   \right]^2
  \right),
\end{equation}
where we introduced $L_2=1/(16e^2)$. This term is nothing but the Skyrme
term. In this way, we end up with the action,
\begin{eqnarray}
 S[U]&=&\frac{F_0^2}{16}\int d^4x 
  {\rm Tr}\left(\partial_\mu U \partial^\mu U^\dagger\right) 
  +\frac{1}{32e^2}\int d^4x
  {\rm Tr}\left(
      \left[
       U^\dagger \partial_\mu U,U^\dagger\partial_\nu U
      \right]^2
     \right)  \nonumber\\
 &&{}+N_c\Gamma[U] 
  +\frac{F_0^2 B_0}{8}\int d^4x
  {\rm Tr}\left(\mathcal{M}^\dagger U+\mathcal{M}U^\dagger\right) 
  \nonumber \\
 &&{}+L_5B_0\int d^4x
  {\rm Tr}\left(\partial_\mu U^\dagger \partial^\mu U
	   \left(\mathcal{M}^\dagger U
	    + U^\dagger \mathcal{M}\right)\right) \nonumber \\
 &&{}+L_8B_0^2\int d^4x
  {\rm Tr}\left(\mathcal{M}^\dagger U\mathcal{M}^\dagger U
	   +\mathcal{M}U^\dagger\mathcal{M}U^\dagger\right),
  \label{action}
\end{eqnarray}
which is up to including $\mathcal{O}(N_c)$ and $\mathcal{O}(p^4)$
terms. Note that there are tree level contributions to $F_\pi$ and
$M_\pi$, and so on. For example,
\begin{equation}
 F_\pi=F_0\left(1+(2m)L_5\frac{16B_0}{F_0^2}\right).
\end{equation}

This action allows a topological soliton, called ``Skyrmion.''  The
classical hedgehog ansatz,
\begin{equation}
 U_c(\bm{x})=
  \left(
   \begin{array}{cc}
    \exp\left(i\bm{\tau}\cdot\hat{\bm{x}}F(r)\right)& 
     \begin{array}{c}
      0\\
      0
     \end{array} \\
    \begin{array}{cc}
     0\ &\  0\\
    \end{array}& 1
   \end{array}
  \right),
\end{equation}
has topological (baryon) number  $B=1$ and stable against
fluctuations. We introduce the collective coordinate $A(t)$,
\begin{equation}
 U(t,\bm{x})=A(t)U_c(\bm{x})A^\dagger(t),
  \label{UUc}
\end{equation}
and treat it as a quantum mechanical degree of freedom.  By substituting
Eq.~(\ref{UUc}) into Eq.~(\ref{action}), we obtain the following quantum
mechanical Lagrangian,
\begin{equation}
 \mathcal{L}=-M_{cl}+\frac{1}{2} \omega^\alpha I_{\alpha\beta}(A)\omega^\beta
  +\frac{N_c}{2\sqrt{3}}\omega^8
  -V(A),
  \label{lag}
\end{equation}
where $\omega^\alpha$ is the ``angular velocity,''
\begin{equation}
 A^\dagger (t)\dot{A}(t)
  =\frac{i}{2}\sum_{\alpha=1}^8 \lambda_\alpha \omega^\alpha(t).
\end{equation}

In the conventional Skyrme model, all the couplings are given in terms
of the $\chi$PT parameters and the integrals involving the profile
function $F(r)$, which is determined by minimizing the classical
energy. In our effective theory approach, on the other hand, they are
determined by fitting the physical quantities calculated by using them
to the experimental values.

The most important feature of the Lagrangian (\ref{lag}) is that the
``inertia tensor'' $I_{\alpha\beta}(A)$ depends on $A$. It has the
following form,
\begin{eqnarray}
 I_{\alpha\beta}(A)&=&I_{\alpha\beta}^0+I'_{\alpha\beta}(A), \\
 I_{\alpha\beta}^0&=&
  \left\{
   \begin{array}{lcl}
    I_1\delta_{\alpha\beta} & \quad & (\alpha,\beta\in\mathcal{I}) \\
    I_2\delta_{\alpha\beta} & \quad & (\alpha,\beta\in\mathcal{J}) \\
    0 & \quad & \mbox{\rm otherwise}
   \end{array}
  \right. \\
 I'_{\alpha\beta}(A)&=&
  \left\{
   \begin{array}{lcl}
    \overline{x}\delta_{\alpha\beta} 
     D_{88}^{(8)}(A)& \quad &(\alpha,\beta\in\mathcal{I})\\
    \overline{y} d_{\alpha\beta\gamma}
     D_{8\gamma}^{(8)}(A) &\quad &
     \begin{array}{l}
      (\alpha\in \mathcal{I},\ \beta\in\mathcal{J}\\
      \mbox{\rm or }\  
       \alpha\in \mathcal{J},\ \beta\in \mathcal{I})\\ 
     \end{array} \\
    \overline{z}\delta_{\alpha\beta}
     D_{88}^{(8)}(A)
     +\overline{w} d_{\alpha\beta\gamma} 
     D_{8\gamma}^{(8)}(A) &  \quad&(\alpha,\beta\in \mathcal{J}) \\ 
    0 & \quad &(\alpha=8\  \mbox{\rm or}\  \beta=8)
   \end{array}
  \right.
\end{eqnarray}
where $\mathcal{I}=\{1,2,3\}$, $\mathcal{J}=\{4,5,6,7\}$, and
$d_{\alpha\beta\gamma}$ is the usual symmetric tensor.

The collective coordinate quantization procedure\cite{Witten:1983tx,Adkins:1983ya,Guadagnini:1983uv,Mazur:1984yf,Jain:1984gp} is
well-known,
and leads to the following Hamiltonian,
\begin{eqnarray}
 H&=&M_{cl}+H_0+H_1+H_2, \\
 H_0&=&\frac{1}{2I_1}\sum_{\alpha\in\mathcal{I}} \left(F_\alpha\right)^2
  +\frac{1}{2I_2}\sum_{\alpha\in\mathcal{J}}\left(F_\alpha\right)^2, \\
 H_1&=&x D_{88}^{(8)}(A)\sum_{\alpha\in\mathcal{I}}\left(F_\alpha\right)^2 
  +y
  \left[
   \sum_{\alpha\in \mathcal{I},\beta\in \mathcal{J}}+
   \sum_{\alpha\in \mathcal{J},\beta\in \mathcal{I}}
  \right]
  \sum_{\gamma=1}^8 
  d_{\alpha\beta\gamma}F_\alpha D_{8\gamma}^{(8)}(A)F_\beta
  \nonumber \\
 &&{}+z\sum_{\alpha \in \mathcal{J}}F_\alpha D_{88}^{(8)}(A) F_\alpha
  +w\sum_{\alpha,\beta \in \mathcal{J}}\sum_{\gamma=1}^8
  d_{\alpha\beta\gamma}F_\alpha D_{8\gamma}^{(8)}(A)F_\beta
  \nonumber \\
 &&{}+\frac{\gamma}{2} \left(1-D^{(8)}_{88}(A)\right), \\
 H_2&=&v\left(
      1-\sum_{\alpha\in \mathcal{I}} \left(D_{8\alpha}^{(8)}(A)\right)^2
      -\left(D_{88}^{(8)}(A)\right)^2
     \right),
\end{eqnarray}
where
\begin{equation}
  x=-\frac{\overline{x}}{2I_1^2},\ \ y=-\frac{\overline{y}}{2I_1I_2},\ \
  z=-\frac{\overline{z}}{2I_2^2},\ \ w=-\frac{\overline{w}}{2I_2^2},
\end{equation}
and $F_\alpha \ (\alpha=1,\cdots, 8)$ are the $SU(3)$ generators,
\begin{equation}
  [F_\alpha,F_\beta]=i\sum_{\gamma=1}^8 f_{\alpha\beta\gamma} F_\gamma,
   \label{F}
\end{equation}
where $f_{\alpha\beta\gamma}$ is the totally anti-symmetric
structure constant of $SU(3)$. Note that they act on $A$ {\em from the
right}.

\section{Fitting the parameters}

We calculate the baryon masses (eigenvalues of the Hamiltonian) in
perturbation theory.  The calculation of the matrix elements of these
operators is a hard task and described in Ref.~\refcite{Harada:2004mj} in detail.
We consider the mixings of representations among
$(\bm{8},\overline{\bm{10}},\bm{27})$ for spin-$\frac{1}{2}$ baryons
and $(\bm{10},\bm{27})$ for spin-$\frac{3}{2}$ baryons.

The best fit set of parameters are obtained by the multidimensional
minimization of the evaluation function, $ \chi^2 = \sum_i
\left(M_i-M_i^{exp}\right)^2/\sigma_i^2$, where $M_i$ stands for the
calculated mass of baryon $i$, and $M_i^{exp}$, the corresponding
experimental value. How accurately the experimental values should be
considered is measured by $\sigma_i$. The sum is taken over the octet
and decuplet baryons, as well as $\Theta^{+}(1540)$ and $\phi(1860)$.
The results are summarized in the following table. 
%%%%%%%%%%%%%%%%%%
 \begin{table}[h]
  {\scriptsize
  \label{expmass}
  \begin{tabular}{c|cccccccccc}
   (MeV)&$\mbox{\rm N}$& $\Sigma$& $\Xi$& $\Lambda$& $\Delta$& ${\Sigma^*}$&
   ${\Xi^*}$& $\Omega$&$\Theta$& $\phi$\\
   \hline
   $M_i^{exp}$ &939 & 1193 & 1318 & 1116 & 1232 & 1385 & 1533 & 1672 & 1539 &
   1862 \\
   $\sigma_i$ &0.6& 4.0 & 3.2 & 0.01 & 2.0 & 2.2 & 1.6 & 0.3 & 1.6 & 2.0
   \\ 
   \hline
   $M_i$ &941 & 1218 & 1355 & 1116 & 1221 & 1396 & 1546 & 1672 & 1547 & 1853
  \end{tabular}
  }
 \end{table}
%%%%%%%%%%%%%%%%%%

The best fit set of values is
\begin{eqnarray}
 M_{cl}&=& 435\mbox{\rm MeV},\ I_1^{-1}=132\mbox{\rm MeV},\ 
  I_2^{-1}=408\mbox{\rm MeV}, \  \gamma=1111\mbox{\rm MeV}, \nonumber \\
 x&=&14.8\mbox{\rm MeV},\ y=-33.5\mbox{\rm MeV},\ z=-292\mbox{\rm MeV},\ 
  w=44.3\mbox{\rm MeV},\nonumber \\
 v&=&-69.8\mbox{\rm MeV},
  \label{bestfit}
\end{eqnarray}
with $\chi^2=3.5\times 10^2$.

Note that they are quite reasonable, though we do not impose any
constraint that the higher order (in $\delta m$) parameters should be
small. The parameter $\gamma$ is unexpectedly large (even though it is
of leading order in $N_c$), but considerably smaller than the value
($\gamma=1573$ MeV) for the case (3) of Yabu and Ando. The parameter $z$
seems also too large and we do not know the reason. Our guess is that
this is because we do not consider the mixings among an enough number of
representations.

\section{Predictions and Discussions}

We have determined our parameters and now ready to calculate other
quantities. First of all, we make a prediction to the masses of the
other members of anti-decuplet,
\begin{equation}
M_{{\rm N}^{\prime}} =1782\ {\rm MeV},
 \quad M_{\Sigma^{\prime}}=1884\ {\rm MeV}.
 \label{massprediction}
\end{equation}
Compare with the chiral quark-soliton model prediction\cite{Ellis:2004uz},
\begin{equation}
 M_{{\rm N}^{\prime}} =1646\ \mbox{\rm MeV},
  \quad M_{\Sigma^{\prime}}=1754\ \mbox{\rm MeV}.
\end{equation}
It is interesting to note that $\Sigma^{\prime}$ is heavier than $\phi$.

The decay widths are such quantities that can be calculated. The results
are reported in Ref.~\refcite{Harada:2004mj}. 

What should we do to improve the results? First of all, we should include
more (arbitrarily many(?)) representations. The mixings with other
representations are quite large, so that we expect large mixings with
the representations we did not include. Second, we may have a better
fitting procedure. In the present method, all of the couplings are
treated equally. The orders of the couplings are not respected. Third,
in order to understand the narrow width of $\Theta^{+}$, we might have
to consider general $N_c$ multiplets\cite{Praszalowicz:2003tc}. Finally
it seems interesting to include ``radial'' modes\cite{Weigel:1998vt}.

\section*{Acknowledgments}

The author would like to thank the organizers for providing this
interesting workshop.

%%%%%%%%%%%%%%%%%%%%%%%%%%%%%%%%%%%%%%%%%%%%%%%%%%%%%%%%%%%%%%%%%%%%

\end{document}